\documentclass[10pt, conference]{IEEEtran}

\usepackage{newcent}
\usepackage{graphicx, color}
\usepackage{amsmath, amssymb}
\usepackage{subfig}
\usepackage{soul}
\usepackage{url}
\usepackage{booktabs}
\usepackage{multirow}
\usepackage{cite}

\begin{document}

\title{Link Capacity Planning for Fault Tolerant Operation in Hybrid SDN/OSPF Networks}

\author{\IEEEauthorblockN{Marcel Caria and Admela Jukan}
\IEEEauthorblockA{Technische Universit\"at Carolo-Wilhelmina zu
Braunschweig\\
m$.$caria@tu-bs$.$de \,\,\,\,\, a$.$jukan@tu-bs$.$de}
}

\maketitle

\begin{abstract}
Link capacity dimensioning is the periodic task where ISPs have to make provisions for sudden traffic bursts and network failures to assure uninterrupted operations. This provision comes in the form of link working capacities with noticeable amounts of \emph{headroom}, i.e., spare capacities that are used in case of congestions or network failures. Distributed routing protocols like OSPF provide convergence after network failures and have proven their reliable operation over decades, but require over-provisioning and headroom of over 50\%. However, SDN has recently been proposed to either replace or work together with OSPF in routing Internet traffic. This paper addresses the question of how to robustly dimension the link capacities in emerging hybrid SDN/OSPF networks. We analyze the networks with various implementations of hybrid SDN/OSPF control planes, and show that our idea of \emph{SDN Partitioning} requires less amounts of spare capacity compared to legacy or other hybrid SDN/OSPF schemes, outperformed only by a full SDN deployment.
\end{abstract}


\section{Introduction}

\par Robust link dimensioning is the periodic task where operators of carrier-class IP networks make provisions for sudden traffic bursts and network failures to assure congestion-free operations. Network operators are required to provision -- in addition to possibly deployed backup capacities in the physical layer -- working capacities with noticeable amounts of \emph{headroom}, i.e., fractions of the capacities that are unused under normal conditions to avoid over-utilization (and thus a degradation of network service quality) in case of a network failure. An essential asset for the calculation of the required headroom is the exact knowledge about the behavior of the control plane in case of sudden events. In the legacy networks this is Open Shortest Path First (OSPF). Today, more and more efforts concentrate on the so-called \emph{hybrid} networking paradigm~\cite{hybrid_1, hybrid_2, hybrid_3, hybrid_4, Brockners, Vissicchio2, tamal_ICC} combining OSPF with Software-Defined Networking (SDN), not the least as it provides advantages of both the legacy- and SDN-based routing.

\par In case of failures, distributed routing protocols, like the legacy OSPF, provide an automatic (autonomous and thus not manipulable) routing convergence and have proven their reliable operation over decades, but require significant over-provisioning of link capacities (e.g., link utilizations of $<50\%$ in the Internet2 backbone~\cite{internet2}). SDN on the other hand provides complete programmability (if desired) that allows for an optimal load-balancing, which in case of failure would require significantly less headroom. Restoration techniques in SDN are to be implemented in the central controller or as an application on top of the controller, like~\cite{ciena-protect}. Therefore, approaches for a hybrid SDN/OSPF control plane appear attractive from the perspective of capacity headroom: they provide the automatic convergence of OSPF, while requiring potentially less link capacity headroom, like in SDN. What the exact benefits of hybrid networking are for fault tolerance, however, has not been studied yet.

\par In this paper, we analyze the problem precisely, and provide a quantitative comparison focused on headroom requirements for various hybrid networks.  We study typical hybrid control planes that follow a "ships-passing-in-the-night'' strategy, whereby legacy routing and SDN control paradigms are oblivious to what the other one configures. We additionally consider our in~\cite{divideandconquer} proposed new scheme for hybrid SDN/OSPF operation called \emph{SDN Partitioning}, where SDN switches are used as invisible border nodes to \emph{partition} the OSPF routing domains. In this scheme, SDN nodes \emph{appear} to their legacy neighbors as regular OSPF routers, while they actually act as simple protocol repeaters that forward all OSPF messages to the centralized SDN controller. Based on a new iterative heuristic for fault-tolerant capacity planning, we show that hybrid networking in general requires less headroom than OSPF. In addition, we show that our idea of \emph{SDN Partitioning} requires the least amount of spare capacity compared to legacy protocols or other hybrid schemes, outperformed only by full SDN deployment.

\par The rest of the paper is organized as follows: Section~\ref{relatedwork} discusses the related work. Section~\ref{arch-section} presents the technological background of hybrid networking. The fault-tolerant capacity planning is detailed in Section~\ref{math-section}. Our numerical study is presented in Section~\ref{results-sect} and Section~\ref{conclusions} concludes the paper.

\begin{figure*}[t] \center
\includegraphics[width=1.0\textwidth]{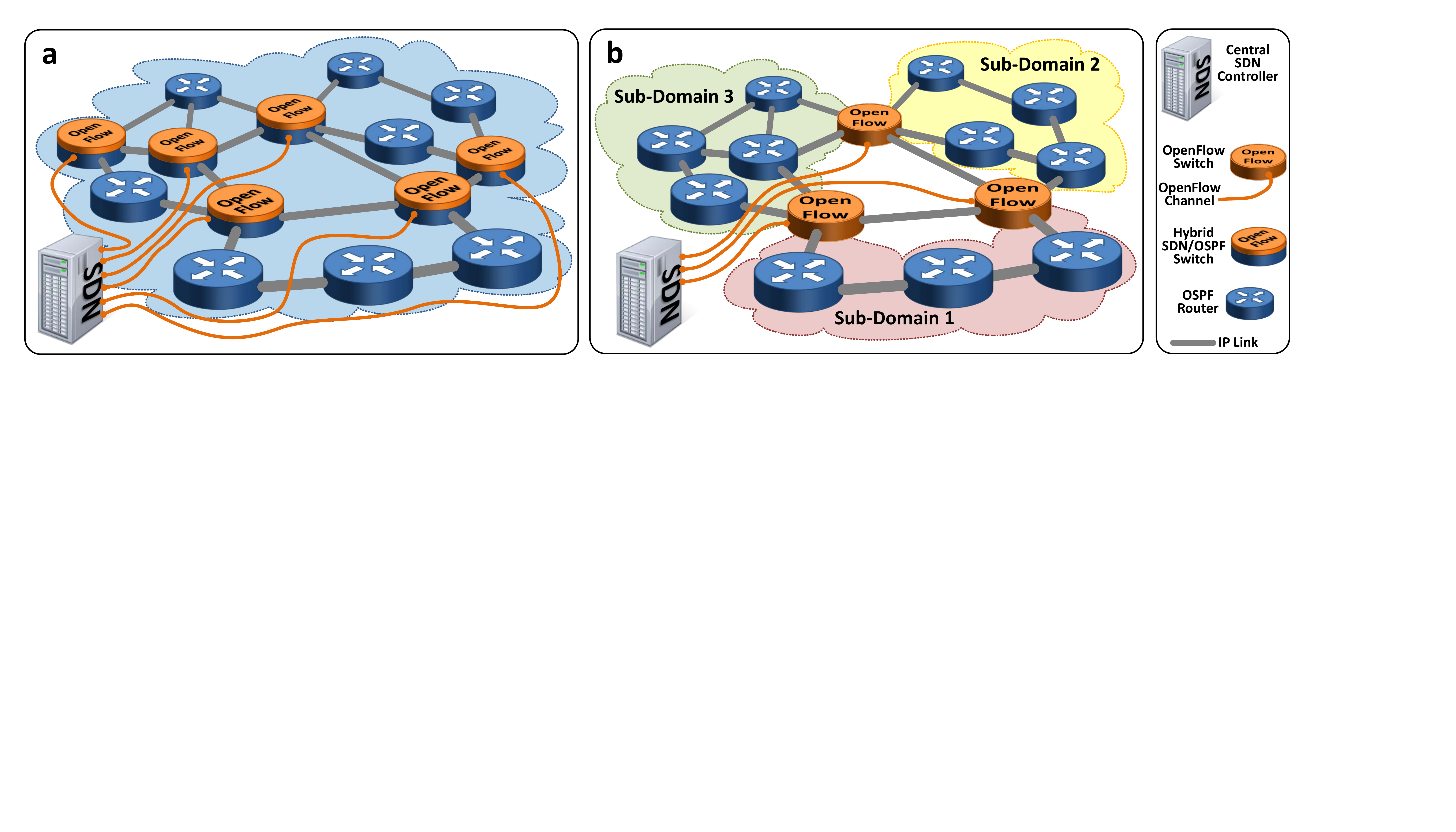}
\caption{The architectures of the two analyzed control planes: a)~\emph{stacked hybrid} and b)~\emph{SDN Partitioning}.}
\label{arch} \end{figure*}

\section{Related Work and Our Contribution}\label{relatedwork}
The problem of determining a sufficient amount of spare capacities considering network failures is not new and has for instance been solved for circuit switched telecommunication backbone networks in~\cite{alrumaih} based on genetic algorithms. In relationship to our approach, the basic principles and best practices of core network capacity planning are explained in~\cite{cisco_cap_planning}. However, hybrid SDN/OSPF networking wasn't considered. It was discussed in~\cite{Kirstaedter} how the increased resilience requirements of new Internet services in IP backbone networks can be met based on MPLS. 

\par Hybrid SDN/OSPF networking has been analyzed and explained to a great extent in~\cite{hybrid_1, hybrid_2, hybrid_3, hybrid_4, Brockners, Vissicchio2, tamal_ICC}, however previous work did not focus on fault tolerance. In a hybrid network, the capability of the SDN controller to insert higher priority rules into the forwarding tables is a powerful new feature which in~\cite{steroids} has  been coined as ``policy based routing on steroids''. We refer to this control plane approach as the \emph{stacked hybrid} model. ``Fibbing'', proposed in~\cite{fibbing1} and~\cite{fibbing2}, shares with SDN Partitioning the idea of steering the legacy routing protocol by introducing \emph{fake} information, but differs significantly from our approach regarding the mode of operation. This paper is the first to extend our previous work on SDN Partitioning~\cite{divideandconquer, caria_HPSR, caria_TNSM} with an analysis of the required link capacities for fault tolerant operation.

\section {Background}\label{arch-section}
OSPF is a distributed routing protocol that requires its local implementation in every router. This provides that all routers in the network can exchange topology information, which in turn allows each router to locally determine for each destination the most suitable port to forward packets by processing a shortest path algorithm. OSPF's information updates are referred to as Link State Advertisements (LSAs) and a router participating in OSPF distributes all its topological information by flooding LSAs throughout the entire network. In case of a network failure, the routers adjacent to that failure start flooding the according topology updates. Each router that receives such an update instantly recomputes its routing and reconfigures its packet forwarding accordingly. This process is referred to as OSPF convergence and may take tens of seconds~\cite{ospf_convergence} in large IP networks.

SDN, in contrast, centralizes (logically) the control over the configuration of the routing in the network, i.e., it separates the control plane from the data plane, which allows to substitute complex routers with dumb forwarding boxes. All SDN switches in the network set up an OpenFlow (i.e., control) channel to the central controller to provide their device and connectivity status information and to receive routing configuration rules. All prominent SDN controller implementations provide some sort of northbound interface to provide access to network management applications, which are used by restoration applications like~\cite{ciena-protect}.

Both control plane mechanisms -- the centralized SDN and the distributed OSPF -- are deployed in a hybrid SDN/OSPF network. Figure~\ref{arch} depicts the two analyzed hybrid control architecture, the stacked hybrid (Figure~\ref{arch}a) and SDN Partitioning (Figure~\ref{arch}b). As it can be seen in both architectures, the actual network includes both OSPF and SDN-enabled (i.e., OpenFlow or hybrid) nodes. SDN-enabled nodes are connected to the central SDN controller. Because OSPF's unimpeded functioning is crucial for hybrid operation, the processing of the LSAs must either be performed at the SDN-enabled nodes, which is required in Figure~\ref{arch}a with all hybrid nodes having OSPF fully implemented, or at the central controller, which is required in Figure~\ref{arch}b. The latter case requires that the SDN controller configures all OpenFlow switches to simply forward (without any processing) all received LSAs to the central controller, which in turn provides it to the network management application that implements the SDN Partitioning scheme and OSPF.

SDN Partitioning, if deployed like in Figure~\ref{arch}b, allows to advertise routing updates individually customized for each of the three sub-domains, which in turn enables to control the routing of traffic flows between these sub-domains. It should be noted that the OSPF routing of sub-domain-internal traffic is not affected at all, but can therefore also not be controlled by the network management application. To enable SDN Partitioning's separation of the OSPF domain into distinct sub-domains, a few SDN-enabled routers must be placed in strategic positions in the network, such that their removal cuts the topology into disconnected components. Obviously, the way the network is clustered into sub-domains is determined solely by the operator's choice of nodes that will be exchanged with SDN nodes, i.e., the sub-domains are determined only once in the beginning and can be changed only by adding new or changing the position of existing SDN nodes. In graph theory parlance, this means that the SDN-enabled routers must constitute a vertex separator. The three example networks used in the numerical evaluation have been partitioned with the algorithm that we introduced in~\cite{caria_TNSM}.

\begin{figure*}[t] \center
\includegraphics[width=1.0\textwidth]{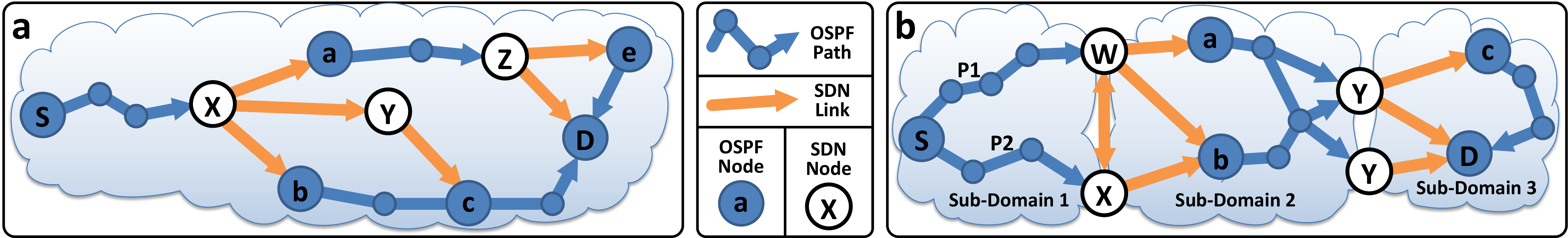}
\caption{The routing capabilities of the two analyzed control planes: a)~\emph{stacked hybrid} and b)~\emph{SDN Partitioning}.}
\label{routing} \end{figure*}


\begin{figure}[t] \center
\includegraphics[width=0.9\columnwidth]{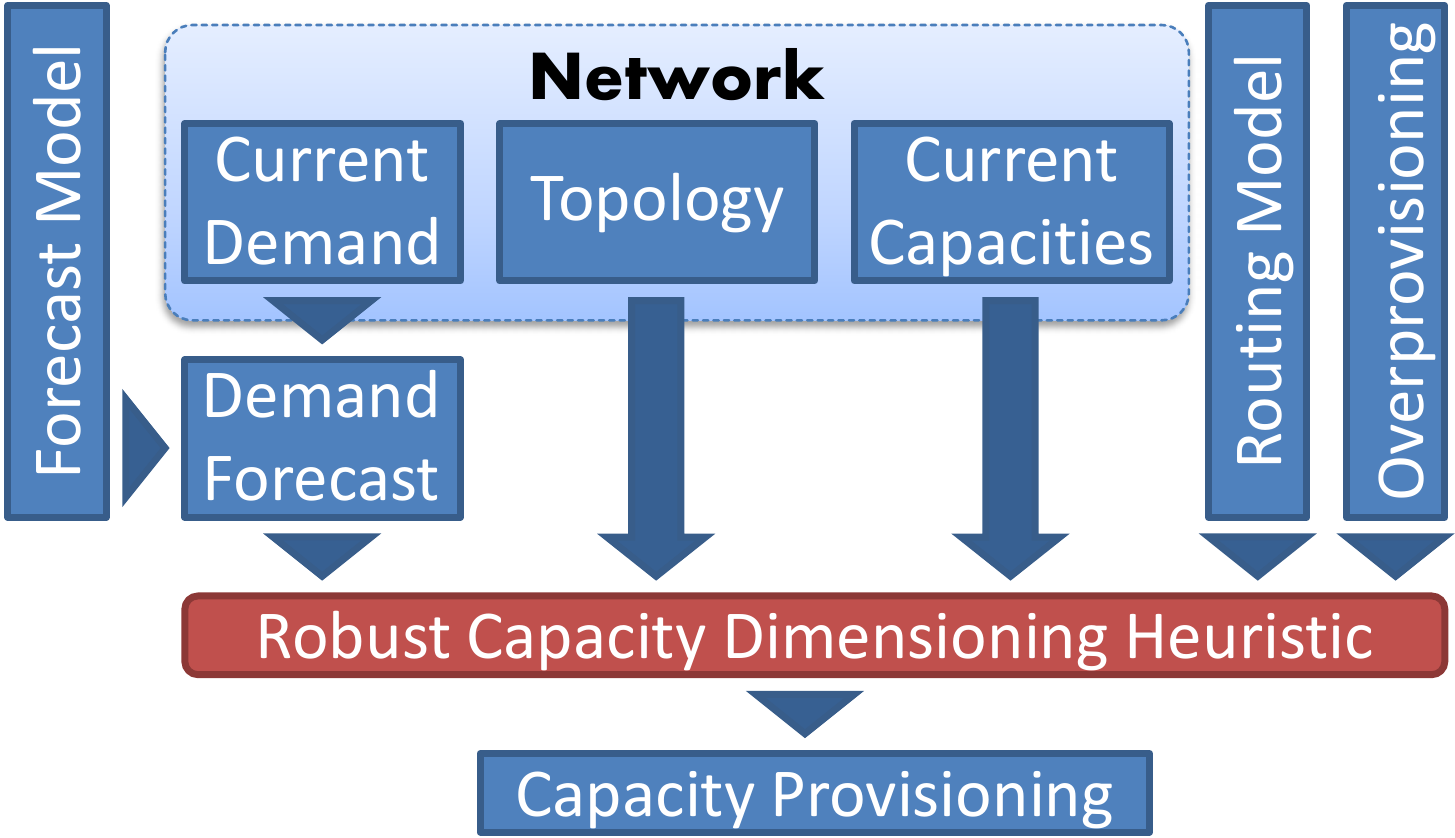}
\caption{The calculation of network capacities and the information it depends on.}
\label{netwPlanning} \end{figure}

\section{Robust Capacity Dimensioning}\label{math-section}
Capacity dimensioning is a periodic network management task, where the links of a network (and the according interfaces of the routers) are dimensioned by the operator for robust operation, i.e., to accommodate the network capacity to the changing traffic demands and events like sudden traffic surges and possible network failures for the next planning period. Like depicted in Figure~\ref{netwPlanning}, capacity dimensioning is a planning process that requires a demand forecast, which extrapolates the demand at the end of the next planning period based on the current demand (which is monitored constantly by the operator) and a forecast model (i.e., statistical methods to estimate the increase of traffic demands based on historical monitoring data). Capacity planning additionally requires topology and capacity information from the network, which is typically available from the network management system in place. Finally, mapping the forecast demand to the links based on the routing model is an optimization process with the objective to minimize the cost of the required capacity upgrades.

\subsection{Traffic Demand Forecasting}
Backbone traffic demand is subject to two kinds of variations relevant for demand forecasting, which are defined by the observed time frame: a)~Daily pattern: Traffic variations on backbone links show a strong daily pattern with typically low utilization in the early morning and maximum utilization during the evening. b)~Annual increase: Traffic demands increase in the long term on average with an annual growth rate that is known to be relatively predictable at least for the next few years. Traffic forecasting has to take both time scales into consideration: The future size of a traffic flow is estimated based on its current daily maximum (a), and upscaled with the annual increase rate (b) till the end of the next planning period. A more sophisticate approach may take flow characteristics of individual demands into consideration, which is however here out of scope, and we assume that we have an exact demand forecast available as input data for the capacity dimensioning process.

\subsection{Overprovisioning} Demand measurements are always averaged over a specific sample interval (e.g., five minute averages). They therefore lack information on the variation \emph{within} each interval that is caused by micro bursts. These bursts can cause short-term congestions, which in turn cause jitter, increased delay, or even packet loss, even though the link may not be highly utilized on average. The relation between the average link load and the required link capacity -- that reduces the frequency of short-term congestions according to the targeted level of network service quality -- is referred to as overprovisioning factor.

\subsection{Routing Model}
The routing model to be used in capacity planning needs to reflect the actual routing configuration capabilities of the network. In case of fixed (i.e., non-dynamic) OSPF link metrics (which is the common case and assumed in our OSPF model), routing changes in an OSPF network are completely predetermined for all network failures, whereas dynamic reactions on sudden traffic changes are impossible. Our OSPF model therefore represents the absolute zero on network programmability. A complete SDN deployment, on the other hand, provides complete freedom regarding the configuration of routing paths, which allows to efficiently load-balance the traffic. SDN therefor represents the full level of network programmability (along with other networking schemes like MPLS or Policy Based Routing).

Hybrid SDN/OSPF operation provides routing configuration capabilities somewhere between these two levels, depending on the number of SDN nodes and their locations, and the used routing model (i.e., stacked hybrid or SDN Partitioning). A valid routing path in such a network is a concatenation of OSPF paths and SDN links, whereas the particular notion of the terms OSPF path and SDN link differs here from common usage: An SDN link is here defined as a directional connection from an SDN router to any other router. An OSPF path is defined as the unique least cost path between an OSPF router and any other (SDN or OSPF) router that doesn't traverse any intermediate SDN node. Figure~\ref{routing} depicts the routing models of the stacked hybrid (\ref{routing}a) scheme and SDN Partitioning (\ref{routing}b) in illustrative example networks. It can be seen in Figure~\ref{routing}a that in a stacked hybrid network, a traffic flow can exit the original OSPF least cost path only when it traverse an SDN node, which can be configured by the central controller to forward the according packets on an arbitrary port.

Figure~\ref{routing}b illustrates SDN Partitioning's additional capability to control routing: The source node $S$ has a distinct least cost path to each SDN border node ($W$ and $X$) in its sub-domain. Thus, the route from $S$ to $D$ starts either with OSPF path $P1$ or $P2$, depending on the aggregated cost metric for the routing to $D$, i.e., the aggregated metrics along $P1$ plus the metric advertised by $W$ for reaching $D$, and the aggregated metrics along $P2$ plus the metric advertised by $X$ for reaching $D$. This example demonstrates that, in contrast to the stacked hybrid scheme, SDN Partitioning provides an additional lever for routing control, which is the metric advertised by the SDN border nodes for each sub-domain external destination.

\begin{figure}[b!] \center
\includegraphics[width=\columnwidth]{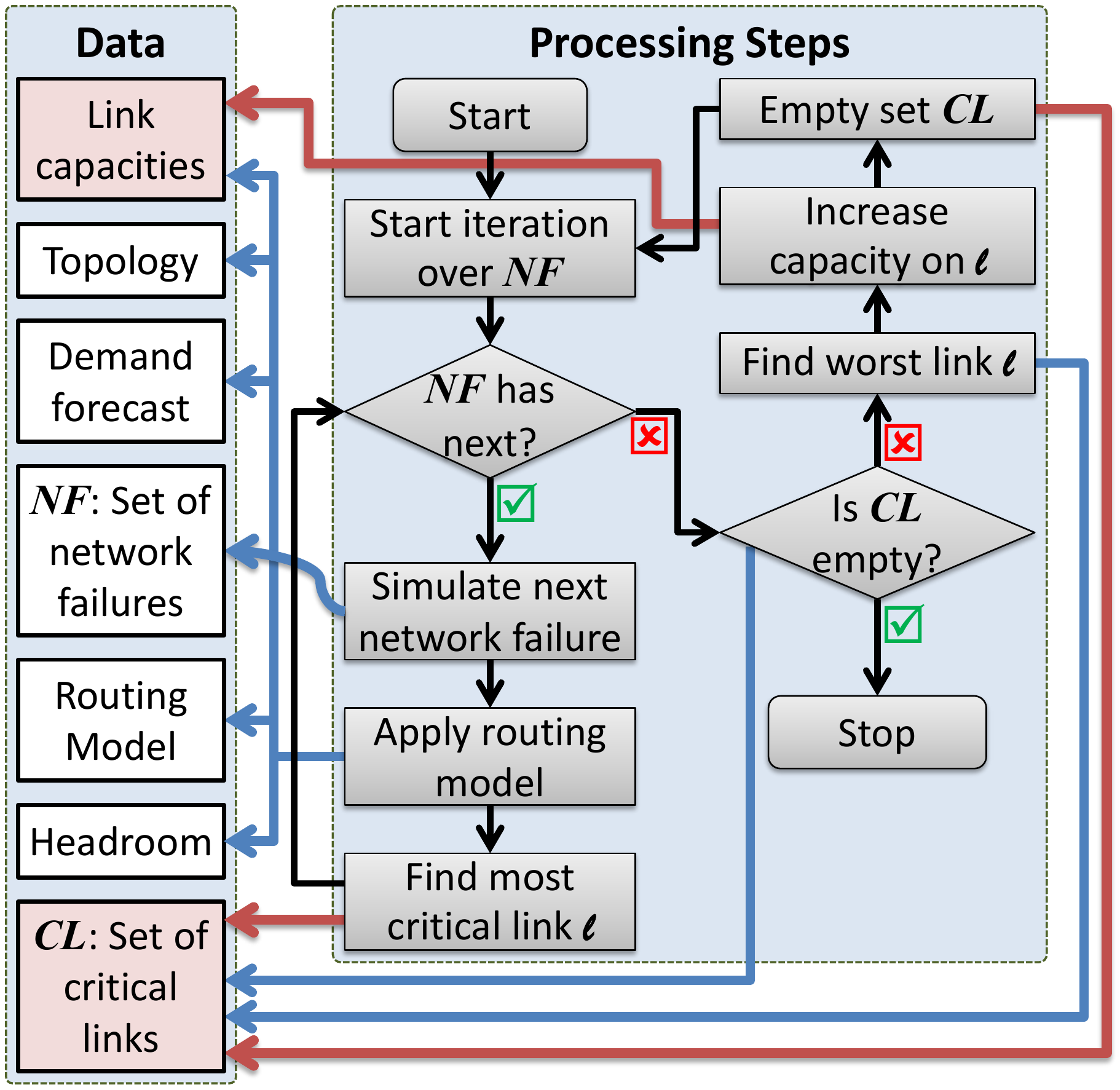}
\caption{The heuristic algorithm for robust link capacity dimensioning.}
\label{algo} \end{figure}

\subsection{Iterative Greedy Algorithm}\label{algo_subsec}
We use a simple heuristic based on an iterative greedy algorithm to determine sufficient link capacities in respect of a predefined set of network failures and the capabilities of the deployed control plane to reroute traffic. The algorithm is depicted as a flow chart in Figure~\ref{algo}, where the boxes in the \emph{Processing Steps} container represent algorithmic states and the boxes in the \emph{Data} container represent working copies of information retrieved from the network management system like explained above. We used black arrows to indicate state transitions of the algorithm, red arrows to indicate write access of a processing step on a data set, and blue arrows to indicate read access. The algorithm consists of two stacked iterations: The inner one iterates over all network failures (depicted as the set $NF$) and applies the routing model to each resulting network scenario (i.e., it tries to load balance the forecast load in the network assuming that the particular failure occurred) to determine the most critical link (which is the one with the largest overload). That link is stored for each failure scenario in the set $CL$. After the inner loop has iterated over all network failures, the algorithm performs one cycle of the outer iteration, where the worst case of all critical links in $CL$ is chosen to be capacity increased (for now only in the working copy of the algorithm) to the next capacity granularity (e.g., from 10~Gbps to 40~Gbps). The algorithm automatically stops when the working copy of the link capacities have been increased to the point where the routing scheme can handle all network failures without overload on any link. The output of the algorithm is the working copy of link capacities that now contains the desired link capacities required for redimensioning the network.

\begin{figure}[t] \center
\includegraphics[width=\columnwidth]{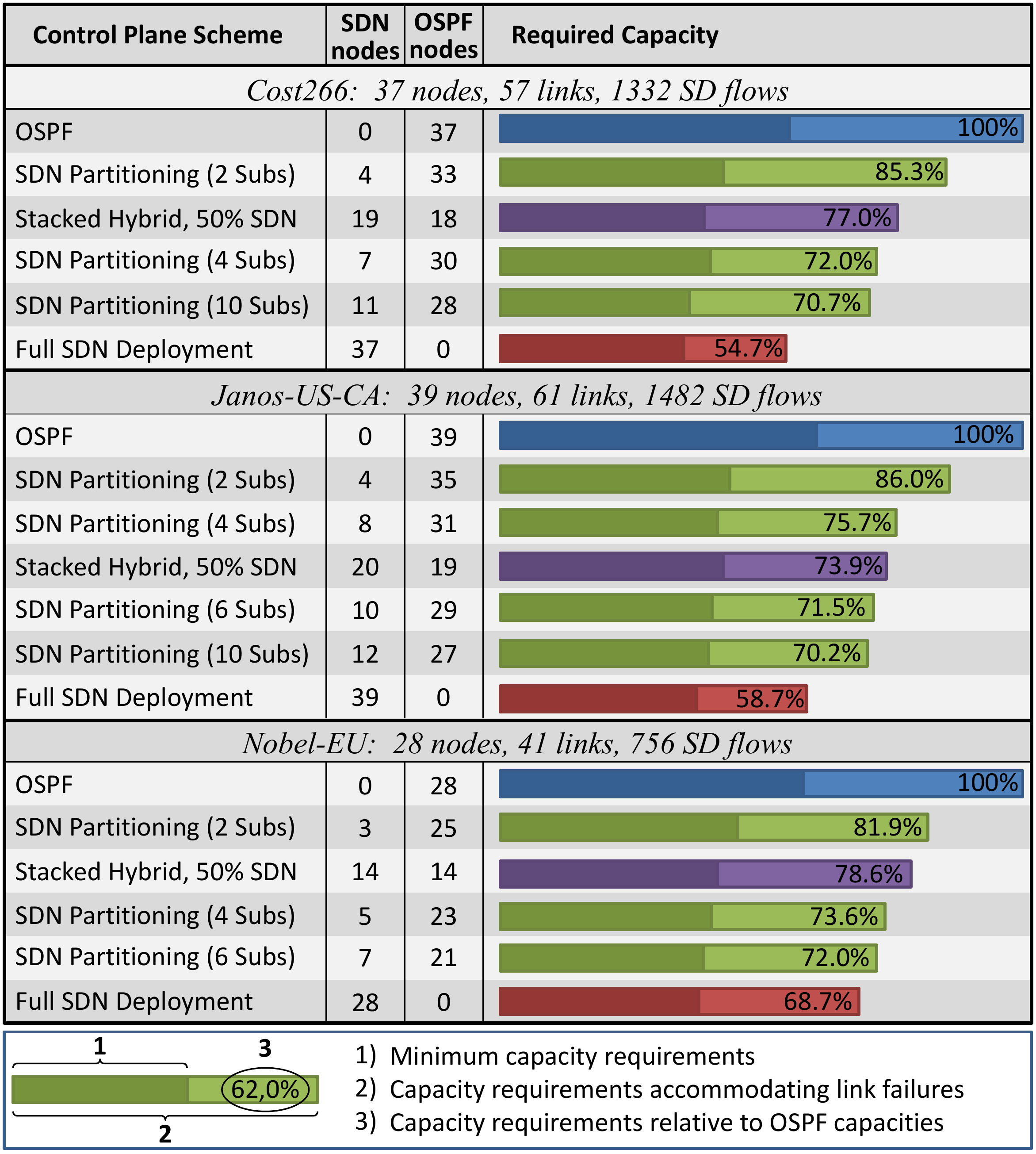}
\caption{Capacity requirements to accommodate link failures for the analyzed routing models.}
\label{result1} \end{figure}

\section{Performance Evaluation}\label{results-sect}
For our performance analysis, we used the Nobel-EU (28 nodes, 41 links), the Cost266 (37 nodes, 57 links), and the Janos-US-CA (39 nodes, 61 links) topologies from the SNDlib library~\cite{sndlib}. Figure~\ref{result1} shows the results of our first experiment, where we used our heuristic for robust link capacity planning. All results are normalized with the capacity requirements of an OSPF-controlled network. Native OSPF is taken as reference scenario, as its routing model provides no load balancing whatsoever, whereas all other operational schemes allow to optimize the routing to improve resource utilization and to reroute traffic more efficiently in case of a failure, which in turn allows to reduce the required link capacities. We compare the capacity requirements of OSPF, full SDN deployment, and the two hybrid control planes: stacked hybrid and SDN Partitioning, whereas the latter is furthermore classified depending on the number of sub-domains in which the initial topology was partitioned. The Cost266 topology was partitioned into 2, 4, and 10 sub-domains, the Janos-US-CA topology was partitioned into 2, 4, 6, and 10 sub-domains, and the Nobel-EU topology was partitioned into 2, 4, and 6 sub-domains.

For the stacked hybrid scheme we assumed that (at least) 50\% of all nodes are SDN-enabled and the optimal locations of these nodes were determined based on the location optimization method in~\cite{hybrid_2}. The actual number of SDN-enabled and legacy OSPF nodes is given in Figure~\ref{result1}. The overlaid (darker) bars show the minimum capacity requirements of a routing scheme without any provisions for network failures. These values have been taken to initialize the capacity planning heuristic (from Subsection~\ref{algo_subsec}). The capacity requirements for fault tolerant operation determined by the heuristic under consideration of all single fiber cuts in the network are shown as the bars in brighter colors, including the relative requirements (in percent) compared to OSPF. Traffic was assumed to be uniformly distributed among all node pairs scaled such that the maximum link load in the OSPF case without link failures was 40~Gbps. Link capacities were available in 10~Gbps, 40~Gbps, and 100~Gbps.

\begin{figure}[b!] \center
\includegraphics[width=1.0\columnwidth]{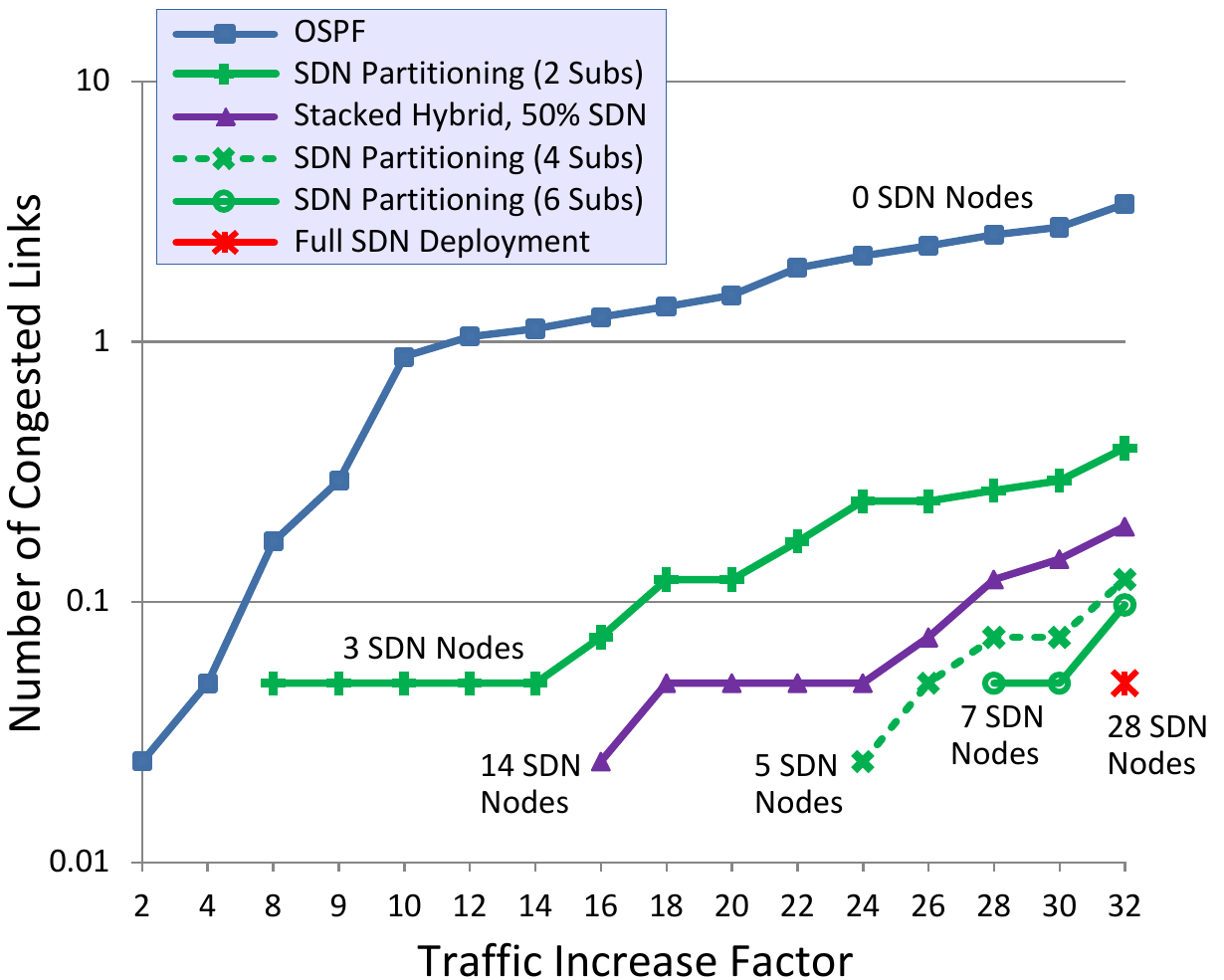}
\caption{Number of congested links in the Nobel-EU topology with a single link failure in case of sudden traffic surges.}
\label{result2} \end{figure}

It can be seen from Figure~\ref{result1} that all hybrid schemes require significantly less capacities than native OSPF for fault tolerant operation, whereas SDN Partitioning requires significantly less SDN-enabled devices to be deployed in the network to achieve results comparable to the stacked hybrid scheme. Even very few SDN nodes operated in SDN Partitioning mode suffice to provide a level of routing control that clearly reduces the capacity requirements. Finally, our results suggest that our method is unsusceptible against the topology of the network, considering the similarities of the results in the three different networks.

Figure~\ref{result2} shows our second experiment, in which we analyzed the behavior of the different routing schemes when sudden traffic surges occur in the coincidental case of a single fiber cut in the network. We here used the Nobel-EU topology with link capacities dimensioned for fault tolerant OSPF operation, and increased the traffic between the two node pairs Madrid - Stockholm and Athens - Glasgow in both directions. The node pairs have been chosen such that they are most distant (geographically and in terms of hop count), thus we stressed the network with four sudden elephant flows with each of them traversing the complete diameter of the network. The original traffic flows between these node pairs were increased with the scaling factors given at the x-axis of Figure~\ref{result2}, and the y-axis (in base-10 log scale) shows the probable number of congested links. The result of this experiment confirms what the previous experiment suggested: The level of routing control in hybrid SDN/OSPF networks provides a significant advantage over native OSPF operation without the investments required for a full SDN deployment. Again, SDN Partitioning outperforms the other hybrid mode with comparably few SDN nodes. It can be seen that the capability of a network operated in native OSPF can not handle elephant flows properly and the probability of congested links in case of a link failure is increasing rapidly with the size of the flows, which suggests that traffic forecasts should rather be upscaled significantly before capacity planning is carried out. Full SDN deployment, on the contrary, appears to be unsusceptible  to traffic surges, as in our experiment it required to scale up the four original flows with factor 32 to see at least any congestion.


\section{Conclusions}\label{conclusions}
We proposed for the first time a method for robust link capacity dimensioning for hybrid SDN/OSPF networks and compared different control plane schemes for a hybrid SDN/OSPF operation. We have detailed the differences of the analyzed hybrid network architectures, and explained the prerequisites, required input data, and functional 
blocks of the proposed capacity planning algorithm. We used the proposed heuristic to compare various control plane schemes regarding their capacity requirements in case of a single link failure in the network and analyzed their capabilities to handle sudden traffic bursts. The results of our numerical evaluation suggest that the requirements of networks operated in legacy OSPF can significantly be reduced be deploying just a few SDN-enabled routers. The low number of required SDN routers especially in case of the SDN Partitioning scheme allows that a relatively high number of nodes can remain in a \emph{configure-once-never-touch-again} operation, which is a known and desired feature of OSPF. Finally, SDN Partitioning showed superior performance in almost all cases of both experiments that were carried out in the course of this work, outperformed only by full SDN deployment, which not only eliminates legacy protocols, which no operator can easily commit to, but also requires significant investments in new networking equipment.



\begin{thebibliography}{1}

\bibitem{hybrid_1} M. Campanella, L. Prete, P.L. Ventre, M. Gerola, E. Salvadori, M. Santuari, S. Salsano, G. Siracusano, \emph{``Bridging OpenFlow/SDN with IP/MPLS,''} poster presentation at TERENA Networking Conference, May 2014, Dublin, Ireland

\bibitem{hybrid_2} M. Caria, A. Jukan, M. Hoffmann, \emph{``A Performance Study of Network Migration to SDN-enabled Traffic Engineering,''} Globecom 2013, Atlanta, USA, December 2013

\bibitem{hybrid_3} D. Levin, M. Canini, S. Schmid, A. Feldmann, \emph{``Incremental SDN Deployment in Enterprise Networks,''} ACM SIGCOMM 2013, Hong Kong, China, August 2013

\bibitem{hybrid_4} S. Agarwal, M. Kodialam, T.V. Lakshman, \emph{``Traffic engineering in software defined networks,''} IEEE INFOCOM 2013, Turin, Italy, April 2013

\bibitem{Brockners} Frank Brockners' Blog at Cisco Blogs: \emph{``Distributed? Centralized? Both?,''} \url{http://blogs.cisco.com/getyourbuildon/distributed-centralized-both}

\bibitem{Vissicchio2} S. Vissicchio, L. Cittadini, O. Bonaventure, G.G. Xie, L. Vanbever, \emph{``On the co-existence of distributed and centralized routing control-planes,''} INFOCOM 2015

\bibitem{tamal_ICC} T. Das, M. Caria, A. Jukan, M. Hoffmann, \emph{``Insights on SDN Migration Trajectory,''} IEEE ICC 2015, London, UK, June 2015

\bibitem{internet2} Internet2 headroom practice: 
\url{http://www.internet2.edu/policies/network-management-policy-interim/}

\bibitem{ciena-protect} Ciena's \emph{Protect,} a centralized global restoration calculator application for SDN,
\url{http://www.ciena.com/connect/blog/Ciena-unveils-Agility-SDNNFV-software-portfolio.html}

\bibitem{divideandconquer} M. Caria, A. Jukan, M. Hoffmann, \emph{``Divide and Conquer: Partitioning OSPF networks with SDN,''} IM 2015, Ottawa, Canada, May 2015

\bibitem{alrumaih} Adel Al-Rumaih, David Tipper, Yu Liu, Bryan A. Norman, \emph{``Spare Capacity Planning for Survivable Mesh Networks,''} Proceedings of the IFIP-TC6 / European Commission International Conference on Broadband Communications, High Performance Networking, and Performance of Communication Networks, Springer, pp.957-968, 12 October 2001

\bibitem{cisco_cap_planning} White Paper, \emph{Best Practices in Core Network Capacity Planning}, Cisco, 2013

\bibitem{Kirstaedter} A. Autenrieth, A. Kirst\"{a}dter, \emph{Fault Tolerance and Resilience Issues in IP based Networks,} Proceedings of the 2nd International Workshop on the Design of Reliable Communication Networks, April 2000

\bibitem{steroids} Brent Salisbury's Blog: \emph{``OpenFlow: Proactive vs Reactive Flows,''} http$:$//networkstatic$.$net/openflow-proactive-vs-reactive-flows/

\bibitem{fibbing1} S. Vissicchio, L. Vanbever, J. Rexford, \emph{``Sweet Little Lies: Fake Topologies for Flexible Routing,''} ACM HotNets 2014, Los Angeles, California

\bibitem{fibbing2} S. Vissicchio, O. Tilmans, L. Vanbever, J. Rexford, \emph{``Central Control Over Distributed Routing,''} ACM SIGCOMM 2015, London

\bibitem{caria_HPSR} M. Caria, A. Jukan, \emph{``The Perfect Match: Optical Bypass and SDN Partitioning,''} HPSR 2015, Budapest, Hungary, July 2015

\bibitem{caria_TNSM} M. Caria, A. Jukan,  M. Hoffmann \emph{``SDN Partitioning: A Centralized Control Plane for Distributed Routing Protocols,''} submitted to IEEE Transactions on Network and Service Management, preprint submitted to arxiv.org

\bibitem{ospf_convergence} M. Goyal, M. Soperi, E. Baccelli, G. Choudhury, A. Shaikh, H. Hosseini, K. Trivedi, \emph{``Improving Convergence Speed and Scalability in OSPF: A Survey,''} IEEE Communications Surveys \& Tutorials, vol.14, iss.2, pp.443-463, May 2012

\bibitem{sndlib} SNDlib library, http$:$//sndlib$.$zib$.$de

\end{thebibliography}
\end{document}